\documentclass{article}
\usepackage{graphicx} 
\usepackage{amsmath}
\usepackage{changepage} 
\usepackage{hyperref}
\usepackage{authblk} 

\title{A statistical note on extending Christensen's limits of agreement with the mean to include subject-observer variation}
\author[1]{Heidi S. Christensen}
\author[1]{Martin Bøgsted}
\author[2]{Jens Borgbjerg}
\affil[1]{Center for Clinical Data Science, Department of Clinical Medicine, Aalborg University, Denmark}
\affil[2]{Department of Radiology, Akershus University Hospital, Norway}
\date{August 2025}

\begin{document}

\maketitle

\section{Introduction}

In Christensen et al.\ we presented a framework for assessing the agreement of continuous measurements made by a number of observers on a number of subjects, extending the limits of agreement with the mean (LOAM) presented by Jones et al.\ to handle more than two observers as well as repeated measurements \cite{Christensen2020, Jones2011}. The LOAM presented by Christensen et al.\ quantifies reproducibility of measurements only and was formulated under a two-way random effects model without including a subject-observer interaction term \cite{Christensen2020}. Here we provide an extension of the framework presented by Christensen et al.\ that allow the assessment of both reproducibility and repeatability as well as assessment of the subject-observer variation. We require repeated measurements and include an interaction term in the two-way random effects model, enabling us to separate residual measurement error from systematic variation in how individual observers measure specific subjects. In addition, repeated measurements allow us to assess both reproducibility, capturing variability due to observers, observer-subject interaction and residual measurement error, and repeatability, reflecting only the residual measurement error. Two measures of LOAM is presented; the \textit{reproducibility LOAM} and the \textit{repeatability LOAM}. The reproducibility LOAM is a measure of how much measurements on the same subject varies across the different observers, while the repeatability LOAM measures how much replicated measurements by an observer on the same subject vary.

\section{Model setup}
Let $Y_{ijk}$ denote the random variable for the measurement made on the $i$th subject by the $j$th observer for the $k$th time, where $i = 1,\dots, a$, $j = 1, \dots, b$, and $k = 1, \dots, c$ for $a, b, c > 1$. In other words, we consider an experiment in which $a$ observers repeatedly assess each of $b$ subjects $k$ times. Then, to define the reproducibility and repeatability LOAM, we assume $Y_{ijk}$ follows a two-way random effect model with interaction given by 
\begin{equation} \label{eq:model}
    Y_{ijk} = \mu + A_i + B_j + (AB)_{ij} + E_{ijk},
\end{equation}
where $A_i \sim N(0, \sigma_A^2)$ for $i = 1, \dots, a$ is the subject effect, $B_j \sim N(0, \sigma_B^2)$ is the observer effect, $(AB)_{ij} \sim N(0, \sigma_{AB}^2)$ is the subject-observer effect, and $E_{ijk} \sim N(0, \sigma_E^2)$ for $k = 1, \dots, c$ is the random measurement error. 

Including the interaction term and requiring repeated measurements ($k > 1$) enables separation of the subject-observer variation from the residual error term, allowing us to assess any systematic differences in how individual observers measure specific subjects by $\sigma_{AB}$. Furthermore, having repeated measurements enables us to quantify not only reproducibility (driven by $\sigma_B^2$, $\sigma_{AB}^2$, and $\sigma_E^2$) but also repeatability (driven solely by $\sigma_E^2)$ as defined in the following.   

\section{The repeatability and reproducibility LOAM}
\subsection{Definition}
Analogous to the LOAM presented in Christensen et al.\ \cite{Christensen2020}, we introduce what we term the \textit{reproducibility LOAM}, defined by

\begin{equation*}
\pm 1.96\sqrt{\text{Var}(Y_{ijk} - \bar{Y}_{i\cdot\cdot})}, 
\end{equation*}
where $\bar{Y}_{i\cdot\cdot}=\frac{1}{bc}\sum_{j,k}Y_{ijk}$ is the average measurement for subject $i$ across all observers and repetitions. Thus, under the two-way random effects model in Equation \eqref{eq:model}, we expect 95\% of the differences to the subject-specific mean, i.e.\, $Y_{ijk} - \bar{Y}_{i\cdot\cdot}$ to fall within these limits. The reproducibility LOAM can also be written as 
\begin{equation}\label{eq:LOAM_reprod}
    \pm 1.96\sqrt{\frac{b-1}{b} (\sigma_B^2 + \sigma_{AB}^2) + \frac{bc-1}{bc}\sigma_E^2}.
\end{equation}
Further, we define the \textit{repeatability LOAM} by 
\begin{equation*} 
\pm 1.96\sqrt{\text{Var}(Y_{ijk} - \bar{Y}_{i j \cdot})}, 
\end{equation*}
where $\bar{Y}_{i j \cdot}$ is the average measurement made on subject $i$ by observer $j$. 
Equivalently we can write the repeatability LOAM as
\begin{equation}\label{eq:LOAM_repeat}
    \pm 1.96 \sqrt{\frac{c -1}{c} \sigma_E^2}.
\end{equation}
The repeatability LOAM provides limits for which we expect 95\% of the differences to the subject-observer-specific mean, i.e., $Y_{ijk} - \bar{Y}_{ij\cdot}$, to be within.

\subsection{Estimation}
Replacing the variance components with their ANOVA estimates (see Table \ref{tab:variance_estimators}) in Equations \eqref{eq:LOAM_reprod} and \eqref{eq:LOAM_repeat}, we obtain an estimate for the reproducibility LOAM given by 
\begin{equation}\label{eq:reprod_est}
    \pm 1.96\sqrt{\frac{b-1}{b} (\hat{\sigma}_B^2 + \hat{\sigma}_{AB}^2) + \frac{bc-1}{bc}\hat{\sigma}_E^2} = 
    \pm 1.96\sqrt{\frac{\text{SSB}+\text{SSAB}+\text{SSE}}{N}},
\end{equation}
where $N = abc$, and similarly, an estimate for the repeatability LOAM is
\begin{equation}\label{eq:repeat_est}
   \pm 1.96 \sqrt{\frac{c -1}{c} \hat{\sigma}_E^2}
 = \pm 1.96\sqrt{\frac{\text{SSE}}{N}}.
\end{equation}

\renewcommand{\arraystretch}{2}
\begin{table}[h]
\centering
\begin{tabular}{|l|}
\hline
ANOVA estimates \\
\hline
$\hat{\sigma}_A^2=\displaystyle \frac{\mathrm{MSA} - \mathrm{MSAB}}{bc}$ \ \\[6pt]
$\hat{\sigma}_B^2=\displaystyle \frac{\mathrm{MSB} - \mathrm{MSAB}}{ac}$ \\[6pt]
$\hat{\sigma}_{AB}^2=\displaystyle \frac{\mathrm{MSAB} - \mathrm{MSE}}{c}$ \\[6pt]
$\hat{\sigma}_E^2 = \mathrm{MSE}$ \\
\hline
\end{tabular}
\caption{ANOVA estimates of the variance components. The mean squares (MSA, MSB, MSAB, and MSE) are defined in Table \ref{tab:anova}.}
\label{tab:variance_estimators}
\end{table}

\renewcommand{\arraystretch}{2}
\begin{table}[h!]
\begin{adjustwidth}{-2cm}{} 
\centering
\begin{tabular}{|c|c|c|c|}
\hline
\textbf{Degrees of freedom} & \textbf{Sum of squares} & \textbf{Mean square} & \textbf{E[MS]} \\
\hline
$\nu_A = a - 1$ 
& $\text{SSA} = bc \sum_i \left( \bar{Y}_{i\cdot\cdot} - \bar{Y}_{\cdot\cdot\cdot} \right)^2$ 
& $\text{MSA} = \frac{\text{SSA}}{\nu_A}$ 
& $bc \sigma_A^2 + c \sigma_{AB}^2 + \sigma_E^2$ \\
\hline

$\nu_B = b - 1$ 
& $\text{SSB} = ac \sum_j \left( \bar{Y}_{\cdot j \cdot} - \bar{Y}_{\cdot\cdot\cdot} \right)^2$ 
& $\text{MSB} = \frac{\text{SSB}}{\nu_B}$ 
& $ac \sigma_B^2 + c \sigma_{AB}^2 + \sigma_E^2$ \\
\hline

$\nu_{AB} = (a - 1)(b - 1)$ 
& \begin{tabular}[c]{@{}l@{}}$\text{SSAB} = c \sum_i \sum_j \left( \bar{Y}_{ij\cdot} - \bar{Y}_{\cdot\cdot\cdot} \right)^2 - \text{SSA} - \text{SSB}$\end{tabular} 
& $\text{MSAB} = \frac{\text{SSAB}}{\nu_{AB}}$ 
& $c \sigma_{AB}^2 + \sigma_E^2$ \\
\hline

$\nu_E = abc - ab$ 
& $\text{SSE} = \sum_i \sum_j \sum_k \left( Y_{ijk} - \bar{Y}_{ij\cdot} \right)^2$ 
& $\text{MSE} = \frac{\text{SSE}}{\nu_E}$ 
& $\sigma_E^2$ \\
\hline
\end{tabular}
\end{adjustwidth}
\caption{ANOVA table for a two-way random effects model with interaction. E[MS] = expected mean square}
\label{tab:anova}
\end{table}

\subsection{Confidence intervals}
As the distribution of the ANOVA estimates for the variance components have no closed form distribution except from $\hat{\sigma}_E$ \cite[p. 132]{Searle1992}, we cannot construct an exact confidence interval for the reproducibility LOAM using the estimator provided in Equation \eqref{eq:reprod_est}. Instead, we provide an approximate confidence interval using the method proposed by Graybill and Wang \cite{Graybill1980}, giving the following approximate 95\% confidence interval for the upper limit of the reproducibility LOAM:
\begin{equation*}
    \left( 1.96 \sqrt{\frac{\text{SSB} + \text{SSAB} + \text{SSE} - L}{N}},\ 
       1.96 \sqrt{\frac{\text{SSB} + \text{SSAB} + \text{SSE} + H}{N}} \right),
\end{equation*}
where 
\begin{align*}
L &= \sqrt{l_B^2 \text{SSB}^2 + l_{AB}^2\text{SSAB}^2 + l_E^2  \text{SSE}^2}, \\
H &= \sqrt{h_B^2 \text{SSB}^2 + h_{AB}^2\text{SSAB}^2 + h_E^2 \text{SSE}^2}
\end{align*}
with $\mathrm{SS}x$ and $\nu_x$ as defined in Table \ref{tab:anova}, $l_x = 1 - \frac{1}{F_{0.975; \nu_x, \infty}}$, and $h_x = \frac{1}{F_{0.025; \nu_x, \infty}} - 1$ for $x = B, \, AB, \,E$. Here $F_{\alpha; m, n}$ is the $\alpha$-quantile of the $F$-distribution with $m$ numerator and $n$ denominator degrees of freedom.
Similarly, an approximate 95\% confidence interval for the lower limit of the reproducibility LOAM is given by
\begin{equation*}
    \left( - 1.96 \sqrt{\frac{\text{SSB} + \text{SSAB} + \text{SSE} + H}{N}}, -1.96 \sqrt{\frac{\text{SSB} + \text{SSAB} + \text{SSE} - L}{N}}\ 
        \right).
\end{equation*}
Since the ANOVA estimate $\hat{\sigma}_E$ follows a $\chi^2$-distribution with $\nu_E$ degrees of freedom (see Table \ref{tab:anova}), an exact 95\% confidence interval for the upper limit of the repeatability LOAM can be constructed based on the estimator in Equation \eqref{eq:repeat_est} and is given by
\begin{equation*}
    \left( 
1.96 \sqrt{ \frac{c - 1}{c} \frac{\text{SSE}}{ \chi^2_{0.975; \nu_E} } },\ 
1.96 \sqrt{ \frac{c - 1}{c} \frac{\text{SSE}}{ \chi^2_{0.025; \nu_E} } } 
\right)
\end{equation*}
with $\nu_E=ab(c-1)$ and SSE as defined in Table \ref{tab:anova} and where $\chi_{\alpha;\nu}^2$ denotes the $\alpha$-quantile for the chi-square distribution with $\nu$ degrees of freedom. Similarly, 
\begin{equation*}
    \left( 
-1.96 \sqrt{ \frac{c - 1}{c} \frac{\text{SSE}}{ \chi^2_{0.025; \nu_E} } } ,\ 
-1.96 \sqrt{ \frac{c - 1}{c} \frac{\text{SSE}}{ \chi^2_{0.975; \nu_E} } }
\right)
\end{equation*}
is an exact 95\% confidence interval for the lower limit of the repeatability LOAM.

\subsection{Sample size calculation based on reproducibility LOAM}
The width of the confidence interval for the reproducibility LOAM is
\begin{equation*}
    W: = \frac{1.96}{\sqrt{N}}
    \left(
    \sqrt{\text{SSB} + \text{SSAB} + \text{SSE} + H } -
    \sqrt{\text{SSB} + \text{SSAB} + \text{SSE} - L }     
    \right).
\end{equation*}
We can write this in terms of the ANOVA estimates of the variance components and the number of subjects, observers, and repetitions by using 
\begin{align*}
    \text{SSB} &= \nu_B \left( a c \hat{\sigma}_B^2 + c \hat{\sigma}_{AB}^2 + \hat{\sigma}_E^2 \right), \\
    \text{SSAB} &= \nu_{AB} \left( c \hat{\sigma}_{AB}^2 + \hat{\sigma}_E^2 \right), \\
    \text{SSE} &= \nu_E \hat{\sigma}_E^2.
\end{align*}
The width of the confidence interval decreases when $H \to 0$ and $L\to 0$; this happens when $h_x \to 0$ and $l_x\to 0$, respectively, for $x = B, AB, E$, which in turn happens when $\nu_x \to \infty$. For fixed $a, c > 1$, all $h_x \to 0$ and $l_x\to 0$ when $b \to \infty$.  To determine the necessary number of observers to achieve a specified width of the confidence interval, we require initial estimates of the relevant variance components, say $\hat{\sigma}_{B,0}^2$, $\hat{\sigma}_{AB,0}^2$, and $\hat{\sigma}_{E,0}^2$ from, for example, a pilot study. Inserting these initial estimates, we obtain
\begin{equation}\label{eq:CI_width}
    W = \frac{1.96}{\sqrt{N}} \left( \sqrt{\text{SSB}_0 + \text{SSAB}_0 + \text{SSE}_0 + H_0} \;-\; \sqrt{\text{SSB}_0 + \text{SSAB}_0 + \text{SSE}_0 - L_0} \right),
\end{equation}
where 
\begin{align*}
H_0 &= \sqrt{h_B^2 \, \text{SSB}_0^2 + h_{AB}^2 \, \text{SSAB}_0^2 + h_E^2 \, \text{SSE}_0^2}, \\
L_0 &= \sqrt{l_B^2 \, \text{SSB}_0^2 + l_{AB}^2 \, \text{SSAB}_0^2 + l_E^2 \, \text{SSE}_0^2},
\end{align*}
and
\begin{align*}
\text{SSB}_0 &= \nu_B \left( a c \hat{\sigma}_{B,0}^2 + c \hat{\sigma}_{AB,0}^2 + \hat{\sigma}_{E,0}^2 \right), \\
\text{SSAB}_0 &= \nu_{AB} \left( c \hat{\sigma}_{AB,0}^2 + \hat{\sigma}_{E,0}^2 \right), \\
\text{SSE}_0 &= \nu_E \hat{\sigma}_{E,0}^2.
\end{align*}
Then, for fixed $a$ and $c$ and a specified width of the confidence interval, we can numerically solve Equation \eqref{eq:CI_width} with respect to $b$ to estimate the number of observers to include to achieve the desired precision. 

\subsection{Comparing two LOAMs}
Often it is of interest to compare LOAM between two methods for measuring a specific entity, for example measuring tumour diameter based on MRI or CT scans. Then, we can calculate the (reproducibility or repeatability) LOAM for each method, say $\mathrm{LOAM}_\mathrm{CT}$ and $\mathrm{LOAM}_\mathrm{MRI}$, and compare the accompanying confidence intervals; non-overlapping confidence intervals indicate a statistically significant difference, while overlapping confidence intervals do not necessarily rule out a significant difference. To formally test for significance, a bootstrap test can be employed as described in the following. 
Consider an experiment where the entity of interest has been measured using both MRI and CT on the same subjects by the same observers and for the same number of repetitions. For illustrative purposes, a toy example with two subjects, two measurements, and two repetitions is shown in Table \ref{tab:tumour_diameters} with measurement values between the two modalities displayed in wide format. Then a bootstrap test can be performed by sampling on the subject-level to preserve the hierarchical structure, that is, for each select subject, all corresponding observations for both measurement methods are included as a whole unit in the bootstrap sample. That is, with the data in Table \ref{tab:tumour_diameters}, whenever we draw a bootstrap sample, we select subjects with replacement from the two available subjects. For each selected subject, we include all eight corresponding rows, that is, all measurements made by the two observers, each with two repeated measurements, for both CT and MRI.

\begin{table}[h!]
\centering
\begin{tabular}{ccccc}
\hline
Subject & Observer & Measurement & CT (mm) & MRI (mm) \\
\hline
1 & 1 & 1 & 26.0 & 25.8 \\
1 & 1 & 2 & 26.2 & 25.8 \\
1 & 2 & 1 & 25.8 & 24.9 \\
1 & 2 & 2 & 25.7 & 24.8 \\
2 & 1 & 1 & 19.0 & 18.2 \\
2 & 1 & 2 & 19.1 & 17.9 \\
2 & 2 & 1 & 19.9 & 19.9 \\
2 & 2 & 2 & 20.1 & 19.7 \\
\hline
\end{tabular}
\caption{Tumour diameter measurements (mm) of two subjects by two observers, each with two repeated measurements, performed using either CT or MRI scans.}
\label{tab:tumour_diameters}
\end{table}

\section{Confidence intervals for variance components}
Using that 
\begin{equation} \label{eq:MS_dist}
    \frac{\nu_X \text{MSX}}{\text{E}[\text{MSX}]}  \sim \chi^2_{\nu_x} \quad \text{for} \, X = B, AB, E
\end{equation}
are independent \cite[p. 132]{Searle1992} (see Table \ref{tab:anova} for definitions of $\nu_x$ and MSX), we provide in the following approximate confidence intervals based on the asymptotic normality of the chi-square distribution for $\sigma_A$, $\sigma_B$, and $\sigma_{AB}$ and an exact confidence interval for $\sigma_E$. 

\subsection{Approximate confidence interval for $\sigma_B$}
By Equation \eqref{eq:MS_dist} and as $\chi^2_\nu \xrightarrow[]{D}N(\nu, 2\nu)$ when $\nu \to \infty$, where $\xrightarrow[]{D}$ denotes convergence in distribution, we have 
\begin{equation*}
\text{MSB} \xrightarrow{D} N \left(
ac\sigma_B^2 + c\sigma_{AB}^2 + \sigma_E^2,\ 
\frac{2}{\nu_B} \left(ac\sigma_B^2 + c\sigma_{AB}^2 + \sigma_E^2 \right)^2
\right)
\end{equation*}
as $b \to \infty$, and
\begin{equation*}
\text{MSAB} \xrightarrow{D} N \left(
c\sigma_{AB}^2 + \sigma_E^2,\ 
\frac{2}{\nu_{AB}} \left(c\sigma_{AB}^2 + \sigma_E^2 \right)^2
\right)
\end{equation*}
as $ab \to \infty$. Then 
\begin{equation*}
\hat{\sigma}_B^2 \xrightarrow{D} N \left(
\sigma_B^2,\ 
\frac{2}{(ac)^2} \left(
\frac{ \left(ac\sigma_B^2 + c\sigma_{AB}^2 + \sigma_E^2 \right)^2 }{ \nu_B } +
\frac{ \left(c\sigma_{AB}^2 + \sigma_E^2 \right)^2 }{ \nu_{AB} }
\right)
\right)
\end{equation*}
as $a,b \to \infty$.
Using plug-in estimates of the variance components, we can construct an approximate 95\% confidence interval for $\sigma_B$ given by

\begin{equation*}
    \hat{\sigma}_B \pm \frac{1.96}{ac \hat{\sigma}_B} \sqrt{
\frac{\left(ac \hat{\sigma}_B^2 + c \hat{\sigma}_{AB}^2 + \hat{\sigma}_E^2 \right)^2}{2\nu_B} +
\frac{\left(c \hat{\sigma}_{AB}^2 + \hat{\sigma}_E^2 \right)^2}{2\nu_{AB}}
}.
\end{equation*}

\subsection{Confidence interval for $\sigma_A$}
By an argument similar to the one for $\sigma_B$, we obtain the following approximate 95\% confidence interval for $\sigma_A$:

\begin{equation*}
\hat{\sigma}_A \pm \frac{1.96}{bc \hat{\sigma}_A} \sqrt{
\frac{\left(bc \hat{\sigma}_A^2 + c \hat{\sigma}_{AB}^2 + \hat{\sigma}_E^2 \right)^2}{2\nu_A} +
\frac{\left(c \hat{\sigma}_{AB}^2 + \hat{\sigma}_E^2 \right)^2}{2\nu_{AB}}
}.
\end{equation*}

\subsection{Confidence interval for $\sigma_{AB}$}

By an argument similar to the one for $\sigma_B$, we obtain the following approximate 95\% confidence interval for $\sigma_{AB}$:
\begin{equation*}
    \hat{\sigma}_{AB} \pm \frac{1.96}{c \hat{\sigma}_{AB}} \sqrt{
\frac{\left(c \hat{\sigma}_{AB}^2 + \hat{\sigma}_E^2 \right)^2}{2\nu_{AB}} +
\frac{\left(\hat{\sigma}_E^2 \right)^2}{2\nu_{E}}
}.
\end{equation*}

\subsection{Confidence interval for $\sigma_E$}
By Equation \eqref{eq:MS_dist}, 
\begin{equation*}
   \left( \frac{\nu_E \text{MSE}}{\chi^2_{0.975;\, \nu_E}}, \, \frac{\nu_E \text{MSE}}{\chi^2_{0.025;\, \nu_E}} \right) 
\end{equation*}
is an exact 95\% confidence interval for $\sigma_E^2$, and in turn 
\begin{equation*}
    \left( \sqrt{ \frac{\nu_E \cdot \text{MSE}}{\chi^2_{0.975,\, \nu_E}} }, \, \sqrt{ \frac{\nu_E \cdot \text{MSE}}{\chi^2_{0.025,\, \nu_E}} } \right)
\end{equation*}
is an exact 95\% confidence interval for $\sigma_E$.

\section{Software}
The R-package \textit{loamr} for calculating LOAMs can be obtained from the GitHub repository \href{https://github.com/CLINDA-AAU/loamr}{https://github.com/CLINDA-AAU/loamr}. In addition to the original (reproducibility) LOAM presented in Christensen et al.\cite{Christensen2020}, this package can also provide both reproducibility and repeatability LOAMs as presented here with the interaction term included in the two-random effects model.  

\bibliographystyle{plain}

\begin{thebibliography}{1}

\bibitem{Christensen2020}
Heidi~S. Christensen, Jens Borgbjerg, Lars Børty, and Martin Bøgsted.
\newblock On jones et al.’s method for extending bland-altman plots to limits of agreement with the mean for multiple observers.
\newblock {\em BMC Medical Research Methodology}, 20(1):304--8, 2020.

\bibitem{Graybill1980}
Franklin~A. Graybill and Chih-Ming Wang.
\newblock Confidence intervals on nonnegative linear combinations of variances.
\newblock {\em Journal of the American Statistical Association}, 75(372):869--873, 1980.

\bibitem{Jones2011}
Mark Jones, Annette Dobson, and Sue O'Brian.
\newblock A graphical method for assessing agreement with the mean between multiple observers using continuous measures.
\newblock {\em International Journal of Epidemiology}, 40(5):1308--1313, 2011.

\bibitem{Searle1992}
Shayle~R. Searle, George. Casella, Charles~E. McCulloch, and Shayle~R. Searle.
\newblock {\em Variance components}.
\newblock Wiley Series in Probability and Statistics. Wiley, New York, 1992.

\end{thebibliography}

\end{document}